# Synthesis and structure of tetragonal $Bi_{12.5}Nd_{1.5}ReO_{24.5}$

N.I. Matskevich*[1,2], Th. Wolf [2], P. Adelmann[2], O.I. Ahyfrieva[1],

M.Yu. Matskevich[1]

[1]*Nikolaev Institute of Inorganic Chemistry, Siberian Branch of the Russian Academy of Sciences, Novosibirsk, 630090, Russia*

[2]*Karlsruhe Institute of Technology, Institute of Solid State Physics, D-76344, Karlsruhe, Germany*

**Abstract**

The $Bi_{12.5}Nd_{1.5}ReO_{24.5}$ tetragonal phase has been synthesized and lattice cell parameters have been determined. According to X-ray data the phase has I4/m symmetry with lattice parameters a = 0.86742 (12) nm, c =1.7408 (3) nm.

**Introduction**

The preparation of new oxide ion conductors with improved characteristics is one of the important task for energy applications. Compounds on the basis of bismuth oxides are new candidate materials for SOFT electrolytes, oxygen ceramic generators, pigments etc [1-10]. They, in particular, have high ionic conductivity in the temperature range of 600-800 K. At present BIMEVOX materials are the most applied. However, the main shortcomings of these compounds are low mechanical stability, high toxicity and easy restoration of vanadium in area with hydrogen high activity. Authors [1, 4-6] discovered new compounds with general formula $Bi_{12.5}R_{1.5}ReO_{24.5}$ (R − rare-earth elements) which have practically the same conductivity as complex oxides on the basis of vanadium and bismuth.

Bismuth oxide has several modifications. δ- $Bi_2O_3$ exists in the temperature range of 1000-1100 K and has cubic structure. In the temperature range of 820-1000 K there is tetragonal modification [7]. Both modifications have high ionic conductivity. The main problem is that modifications exist in narrow temperature range. There are many attempts to stabilize these modifications up to room temperatures in literature. In papers [1, 4-6] synthesis and structural investigations of several compounds with composition $Bi_{12.5}R_{1.5}ReO_{24.5}$ (R − rare-earth elements) were performed. As was shown the phases had cubic structures. Compounds with lanthanum and neodymium ($Bi_{12.5}Nd_{1.5}ReO_{24.5}$, $Bi_{12.5}La_{1.5}ReO_{24.5}$) possess highest ionic conductivity. The substitution of $Bi_2O_3$ by rhenium allows one to stabilize phase using even largest lanthanides.

In our paper synthesis of new tetragonal modification $Bi_{12.5}Nd_{1.5}ReO_{24.5}$ and its lattice parameters are presented.

**Experimental part**

Preparation methods of mixed oxides on the basis of bismuth and rare-earth elements can be classified as: solid-phase synthesis, crystals growth, synthesis in solution. Usually the method of ceramic technology is applied to prepare polycrystalline materials in $Bi_2O$-$R_2O_3$-$Re_2O_7$ system. Synthesis is performed as a rule from bismuth oxide, lanthanide oxides and rhenium oxide or perrhenate ammonium at high temperature. The method includes usually following technological operations: heat treatment of initial substances, their mixing, pressing, and agglomeration. These operations repeat many times. Temperatures vary from 800 up to 1100 K.

There are two ways of $Bi_{12.5}R_{1.5}ReO_{24.5}$ synthesis in literature: 1) preparation from bismuth oxide, rare-earth oxides and rhenium oxide; 2) synthesis from $Bi_2O_3$, $R_2O_3$ and $NH_4ReO_4$ [1, 4-6]. Characterization is carried out by methods of X-ray phase analysis.

In our paper the synthesis was performed according to reaction: $6.25Bi_2O_3$ + $NH_4ReO_4$ + $0.75Nd_2O_3$ = $Bi_{12.5}Nd_{1.5}ReO_{24.5}$ + $NH_3$ + $0.5H_2O$. The following rea-gents were used for preparation: $Bi_2O_3$ (99.999%, ABCR), $NH_4ReO_4$(>99%, Alfa Aesar, Johnson Matthey Company), $Nd_2O_3$ (99.9%, Reacton, Rare Earth Products, Division of Johnson Mattey Chemical LTD).

The samples were prepared as follows. Starting reagents were treated before synthesis at 1100 K ($Nd_2O_3$) up to constant weight. Then they were mixed in an agate mortar and ground for several hours with several intermediate reground, pressed and heat treated in air at 800-900 K. The phase purity was analyzed with X-ray diffractometer (STADI-P, Stoe diffractometer, Germany, Cu $K_{\alpha 1}$ radiation). X-ray analysis showed that samples were single phase.

**Results and discussion**

X-ray diffraction data were used to calculate the lattice parameters of the new compound: low temperature $Bi_{12.5}Nd_{1.5}ReO_{24.5}$. The observed X-ray pattern of $Bi_{12.5}Nd_{1.5}ReO_{24.5}$ is presented in Figure 1.

This phase crystallizes in a tetragonal structure (space group I 4/m) with a refined cell parameters: a = 0.86742 (12) nm, c = 1.7408 (3) nm, V = 1.3098 (4) $nm^3$. Detailed information is presented in Table 1.

**Conclusions**

The low temperature (tetragonal) modification of $Bi_{12.5}Nd_{1.5}ReO_{24.5}$ has been prepared for the first time. It has been shown that phase has tetragonal symmetry (space group I4/m) with lattice parameters a = 0.86742 (12) nm, c =1.7408 (3) nm.

**Acknowledges**

This work was supported by Karlsruhe Institute of Technology, RFBR (Project 13-08-00169) and Program of Fundamental Investigation of Siberian Branch of the Russian Academy of Science.

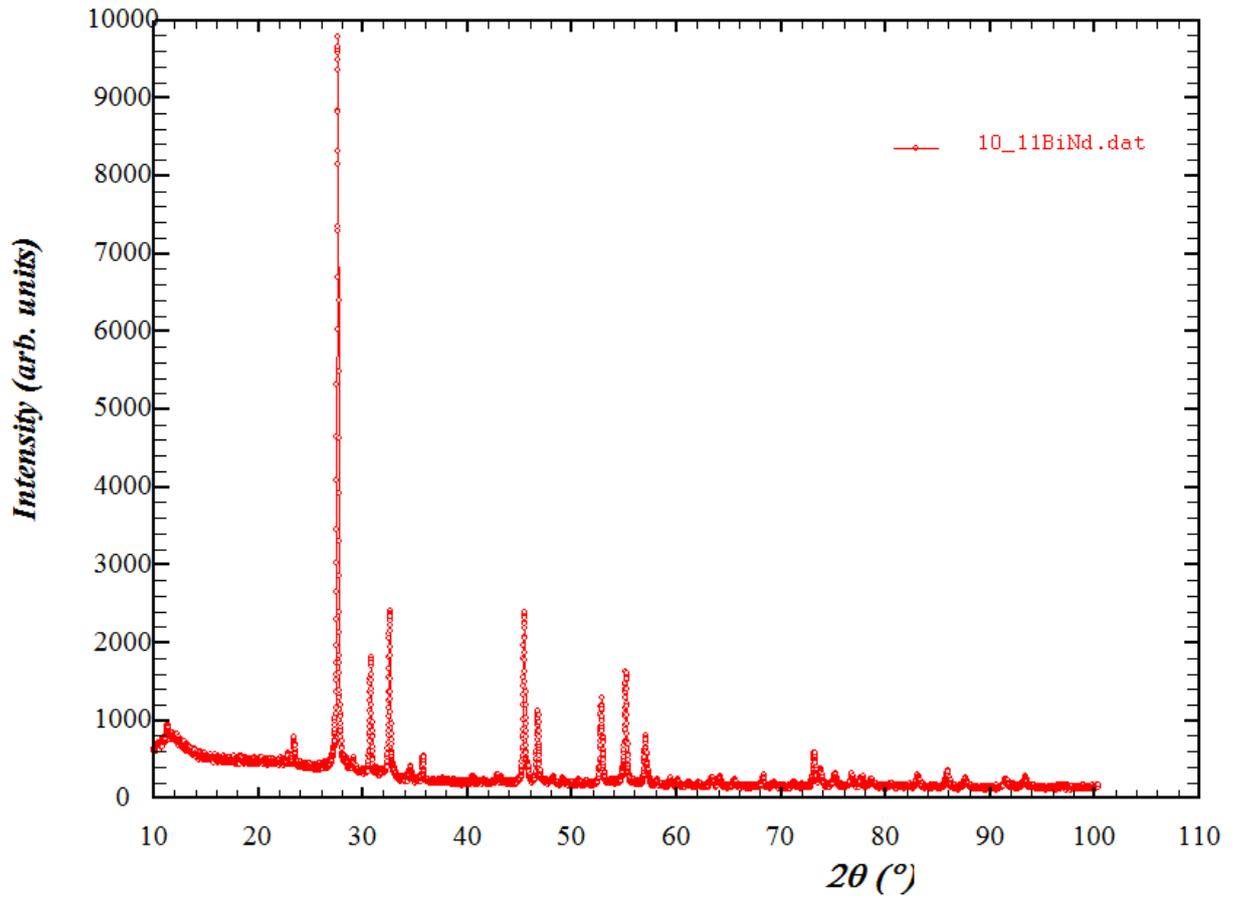

Figure 1. X-ray power diffraction of low temperature phase $Bi_{12.5}Nd_{1.5}ReO_{24.5}$

Table 1. Power X-ray diffraction data for tetragonal $Bi_{12.5}Nd_{1.5}ReO_{24.5}$

| N | 2Th [obs] | H | K | L | 2Th[calc] | obs-calc | Int. | d[obs] | d[calc] |
|---|---|---|---|---|---|---|---|---|---|
| 1 | 11.368 | 1 | 0 | 1 | 11.388 | -0.0204 | 9.5 | 7.7776 | 7.7637 |
| 2 | 22.878 | 2 | 0 | 2 | 22.891 | -0.0127 | 5.8 | 3.8840 | 3.8819 |
| 3 | 23.466 | 2 | 1 | 1 | 23.477 | -0.0104 | 7.4 | 3.7880 | 3.7863 |
| 4 | 27.640 | 2 | 1 | 3 | 27.638 | 0.0019 | 100.0 | 3.2247 | 3.2249 |
| 5 | 30.812 | 0 | 0 | 6 | 30.794 | 0.0183 | 18.0 | 2.8996 | 2.9013 |
| 6 | 32.616 | 3 | 1 | 0 | 32.618 | -0.0029 | 24.4 | 2.7433 | 2.7430 |
| 7 | 34.599 | 2 | 1 | 5 | 34.590 | 0.0089 | 4.2 | 2.5904 | 2.5910 |
|   |   | 3 | 0 | 3 | 34.633 | -0.0344 |   |   | 2.5879 |
| 8 | 35.791 | 2 | 2 | 4 | 35.790 | 0.0009 | 5.2 | 2.5068 | 2.5069 |
| 9 | 42.911 | 2 | 2 | 6 | 42.875 | 0.0363 | 3.1 | 2.1059 | 2.1076 |
|   |   | 4 | 0 | 2 | 42.947 | -0.0359 |   |   | 2.1042 |
| 10 | 45.479 | 3 | 1 | 6 | 45.469 | 0.0102 | 23.9 | 1.9928 | 1.9932 |
| 11 | 46.798 | 4 | 0 | 4 | 46.765 | 0.0323 | 11.1 | 1.9397 | 1.9409 |
|   |   | 4 | 2 | 0 | 46.799 | -0.0015 |   |   | 1.9396 |
| 12 | 49.186 | 3 | 3 | 4 | 49.198 | -0.0124 | 2.6 | 1.8509 | 1.8505 |
| 13 | 50.667 | 4 | 1 | 5 | 50.656 | 0.0107 | 2.4 | 1.8002 | 1.8006 |
| 14 | 52.859 | 2 | 1 | 9 | 52.848 | 0.0105 | 12.5 | 1.7306 | 1.7310 |
| 15 | 55.231 | 5 | 0 | 3 | 55.218 | 0.0131 | 16.2 | 1.6618 | 1.6621 |
| 16 | 57.083 | 4 | 2 | 6 | 57.073 | 0.0104 | 8.0 | 1.6122 | 1.6125 |
| 17 | 60.196 | 4 | 0 | 8 | 60.198 | -0.0015 | 2.4 | 1.5360 | 1.5360 |
| 18 | 63.408 | 5 | 3 | 2 | 63.379 | 0.0289 | 2.5 | 1.4657 | 1.4663 |
|   |   | 2 | 1 | 11 | 63.431 | -0.0225 |   |   | 1.4653 |
| 19 | 64.152 | 0 | 0 | 12 | 64.147 | 0.0050 | 2.7 | 1.4505 | 1.4506 |
| 20 | 65.582 | 5 | 0 | 7 | 65.556 | 0.0258 | 2.3 | 1.4223 | 1.4228 |
| 21 | 68.331 | 6 | 0 | 4 | 68.313 | 0.0188 | 2.8 | 1.3716 | 1.3720 |
|   |   | 6 | 2 | 0 | 68.339 | -0.0077 |   |   | 1.3715 |
| 22 | 73.235 | 5 | 0 | 9 | 73.233 | 0.0023 | 5.9 | 1.2914 | 1.2915 |
| 23 | 73.811 | 3 | 1 | 12 | 73.839 | -0.0277 | 3.7 | 1.2828 | 1.2824 |
| 24 | 75.220 | 5 | 4 | 5 | 75.201 | 0.0191 | 3.2 | 1.2622 | 1.2625 |
|   |   | 6 | 3 | 3 | 75.226 | -0.0063 |   |   | 1.2621 |
| 25 | 76.824 | 6 | 2 | 6 | 76.813 | 0.0114 | 3.0 | 1.2398 | 1.2399 |
| 26 | 77.780 | 7 | 1 | 0 | 77.796 | -0.0156 | 2.7 | 1.2269 | 1.2267 |
| 27 | 83.067 | 4 | 2 | 12 | 83.072 | -0.0048 | 2.9 | 1.1617 | 1.1617 |
| 28 | 85.958 | 7 | 1 | 6 | 85.963 | -0.0050 | 3.4 | 1.1299 | 1.1299 |
|   |   | 4 | 1 | 13 | 85.984 | -0.0265 |   |   | 1.1296 |
| 29 | 87.662 | 6 | 0 | 10 | 87.674 | -0.0121 | 2.5 | 1.1123 | 1.1122 |